\documentclass[runningheads]{llncs}

\usepackage{graphicx}
\usepackage{epsfig}
\def\esp{\hspace*{.2in}}

\begin{document}

\title{Artificial Agents and Speculative Bubbles}
\author{Y. Semet, S. Gelly, M. Schoenauer, M. Sebag}
\institute{Equipe TAO, Universit\'e d'Orsay, CNRS, INRIA, 91405 Orsay, France.}

\date{~}

\maketitle


\begin{abstract}
Pertaining to Agent-based Computational Economics (ACE), this work presents two
models for the rise and downfall of speculative bubbles through an exchange
price fixing based on double auction mechanisms. The first model is based on a
finite time horizon context, where the expected dividends decrease along time.
The second model follows the {\em greater fool} hypothesis; the agent behaviour
depends on the comparison of the estimated risk with the greater fool's.
Simulations shed some light on the influent parameters and the necessary
conditions for the apparition of speculative bubbles in an asset market within
the considered framework.
\end{abstract}


\section{Introduction}

Beyond the standard economics models, traditionally centered on Rational
Expectations Equilibria \cite{Radner:79} and efficient markets, several new
paths have recently been explored to tackle the otherwise unexplainable
phenomena underlying speculative bubbles. Having an idea of what is behind
these phenomena is crucial as they question the ability of existing markets to
perform efficient resource allocation. Their understanding is a necessary first
step on the way leading to the design of safer market structures.
????\cite{Grossman:80}????

One approach for investigating speculative bubbles is Experimental Economics,
as pioneered by Vernon Smith; several studies \cite{Smith:91,Noussair:01}
demonstrate the rise and downfall of speculative bubbles in closed and
controllable laboratory environments involving candid and/or experimented human
beings. Such experiments also offer room for studying the effects of
information cost and/or sociological noise (or cognitive dissonances).

Another approach is Agent-based Computational Economics (ACE; see
\cite{Tesfatsion:01a} for a survey). This bottom-up approach, based on
Artificial Intelligence (AI)-oriented models of economic agents, replaces
experiments by computational simulations. It also offers a controllable
framework for studying the emergence and dynamics of global patterns from the
repeated interaction of elementary agents endowed with limited perception,
communication, cognitive and learning abilities. Financial markets have been
studied intensively along these lines (see
\cite{LeBaron:00,LeBaron:01a,Semet:2003d}), and some inspirations for the
present work will be discussed in Section \ref{StateofArt}.

Pertaining to the ACE field, this paper investigates two models for boundedly
rational agents, inspired from Duffy and Unver's work on goods markets
\cite{Duffy:2003}. Likewise, we consider Stochastic-Zero-Intelligence  (SZI)
agents, whose bids and asks are randomly drawn from a price distribution
depending on the previous exchange price. The difference lies in the
distribution setting strategy chosen to explain speculative bubbles.

Two strategies are enforced 
with respectively finite and infinite time horizons. In the first setting
(exogenous risk), the risk straightforwardly depends on the expected dividends.
In the second setting (endogenous risk), the agent strategy is determined from
the comparison between the estimated risk, and the agent's and {\em greater
fool}'s risk thresholds. Depending on this comparison, the agent's strategy is
exuberant, comfortable, or panicky, thereby ruling its propensity to buy or
sell ($pbs$) and its bid and ask distribution.

The paper is organized as follows. After briefly reviewing some related works
(Section \ref{StateofArt}), we describe the market mechanism and the agent models
(Section \ref{models}). Section \ref{Expe} reports on experimental simulation
results, and discusses the necessary conditions and influent parameters with
respect to bubble dynamics. The paper ends with perspectives for further
research.

\section{Related works} \label{StateofArt}

Let us briefly review some sources of inspiration for the present work.
Beltratti and Margarita \cite{Beltratti:92} consider an artificial market where
individual agents maximize their expected return, measured after an individual
price estimation mechanism. This estimation is achieved through artificial
neuron networks. The cost of information is accounted for as the number of
neurons in the NN's hidden layer. As agents might decide to invest in a simple,
average or complex price estimate, one observes the general market behaviour,
the distribution of simple, complex and other agents, and the individual
strategies pay off.

Arifovic \cite{Arifovic:96} proposes a 3-parameters agent model, governing the
exchange rate between two currencies. The three parameters are evolved along a
simple Genetic Algorithm. One major interest of this work is to reproduce the
market behaviour observed in experimental economics (oscillations), contrasting
with the dynamics predicted by rational expectation theory.

In the famous ``El Farol'' problem \cite{Arthur:94}, another dimension for
bounded rationality appears, namely the anticipation of other agents decisions.
Each agent will decide to go to the bar, if and only if it expects the bar to
be reasonably crowded (follow the minority rule). Along the same lines, the
Santa Fe artificial stock market \cite{Arthur:97,LeBaron:99,LeBaron:02}
provides a unified framework where agents are endowed with forecasting rules
(evolutionary classifier systems). Depending on the evolution pace (the rules
refreshing, or the information cost), the  behaviour switches from an efficient
to a speculative market.

\section{Overview}\label{models}

The market considered in the following involves a finite set of agents, trading
a single asset.

\subsection{Exchange rule}

The exchange rule is based on a {\em double auction} mechanism (Table
\ref{algo:marketClearing}). Each novel ask (respectively bid) is compared with
the current selling (resp. buying) order book; it succeeds whether it is
greater (resp. lower) than the current minimum selling order (resp. the current
maximum buying order), and the order is then removed from the book; otherwise,
the order books are updated with the novel ask (resp. bid) offer. Both books
contain at most one offer from each agent ({\em cleared book} convention).

\begin{table}
\centerline{\begin{tabular}{l}
{\bf Init}: Initialize agents; Buying order book = $\{ \}$; Selling order book = $\{ \}$; \\
{\bf Loop}: For each auction round, time $t$\\
\esp\esp For each agent \hfill {\em random permutation}\\
\esp\esp\esp $Order(bid \vee ask \vee idle; value)$ = strategy(agent)\\
\esp\esp\esp If $Order$ succeeds\\
\esp\esp\esp\esp Exchange price = $P_{currentBest}$\\
\esp\esp\esp\esp Refresh order book\\
\esp\esp\esp Else\\
\esp\esp\esp\esp Update ($Order$, order Books)\\
\esp\esp $p_t$ = Average Exchange price over the round\\
\end{tabular}}
\caption{Market Clearing Algorithm} \label{algo:marketClearing}
\end{table}

The agent order is determined according to the agent strategy detailed below;
the exchange price is set to the current best offer (the minimum selling order
on an ask and the maximum buying order on a bid).

\subsection{Individual agents}
    Agent $i$ is characterized from its belongings, cash and number of
shares ($cash(i,t)$ and $shares(i,t)$) and its estimation of the asset
fundamental value ($F(i,t)$).

In each auction round, the agent decides between buying one share, selling one
share, or remaining idle. The choice depends on its strategy, detailed below,
and based on its estimation of the risk currently held by the market.

The strategy governs the decision and the price offer. In summary, the bounded
rationality is made up three elements:
    \begin{itemize}
        \item A risk estimation function.
        \item A strategy that maps risk into a decision (buy $\vee$ sell $\vee$ idle) and a price offer.
    \end{itemize}

\subsubsection{Exogenous risk and finite time horizon}

A straightforward strategy is based on the finite time horizon $t = 0..T$: the
propensity to buy of each agent decreases as the game goes to an end. The risk
linearly increases from 0 at time $t=0$ up to 1 at $t = T$. The propensity to
buy decreases as the risk increases. Several models have been considered
(linear, sigmoid, exponential), see Section \ref{Expe}.

The pricing strategy, inspired from the anchoring effect \cite{Duffy:2003},
follows a uniform distribution centered on the previous exchange price
$p(t-1)$. If $U_{[a,b]}$ stands for the uniform distribution on segment
$[a,b]$, \[ Order(t ; buy \vee sell; level): \esp p(buy) \propto \frac{t}{T};
\esp level \propto p(t-1)+U_{[-1\%;+1\%]} \]

Clearly, this model suffers from two shortcomings. On one hand, although finite
time horizons are consistent with experimental economic settings (e.g.
\cite{Smith:91,Ruffieux:03,Noussair:01}), they are not with respect to actual
markets. Second, this model offers limited insights into the causes of
speculative bubbles as the market behaviour is ultimately controlled from the
(exogenous) risk function (as $p(buy) = f(\frac{t}{T})$, a variety of price
curves can be obtained through carving function $f$).

\subsubsection{Endogenous risk}

To get rid of the aforementioned limitations, a more sophisticated bounded
rationality model is proposed. This model involves a naive form of technical
trading: the risk estimation and subsequent decisions are based on internal
parameters (among which the agent's estimate of the asset fundamental value,
$F(i,t)$), and the exchange price history $p(t)$.

More specifically, risk is computed from two terms: the distance between the
current price and the asset fundamental value (agent internal parameters), and
the slope of the price curve (averaged on the 3 previous time steps).

The first term accounts for potential arbitrage profits: risk increases as the
exchange price gets higher than the asset fundamental value. The second term
reflects the ``greater fool'' hypothesis: as the price wildly increases,  the
risk actually decreases as a greater fool is likely to buy your shares.

The weighted sum of the two above terms is taken through a sigmoid, ensuring
that the risk estimate varies smoothly in $[0,1]$. Finally, where $v_i$ and
$w_i$ denote the weights (agent internal parameters) for the deviation from
fundamental value and the price slope,

\[
  r (i,t) = \frac{1}{1 + a e^{-(v_i(p(t)-F(i,t))-w_i \frac{dp}{dt}}} \]

The sigmoid's slope (factor $a$) controls the transition between the low and
high risk regions.

\subsubsection{Exuberance, comfort and panic strategies}

The agent risk is compared to two thresholds: the agent risk threshold $R_i$
(internal parameter) and the fool's threshold (set to $\alpha_i \times R_i$).

The comparison determines the agent strategy:

\begin{description}
\item[Exuberant] The risk is smaller than the agent risk threshold ($r(i,t) < R_i$).\\
In this case, the agent tends to buy a share (with probability 80\%); otherwise, it stays idle or sell a share with equal probability (10\%).\\
The price offer is drawn from the uniform distribution centered on the previous
exchange price $p(t-1)+U_{[-1\%;+1\%]}$.

\item[Comfort] The risk is between the agent risk threshold and the fool's
threshold ($R_i < r(i,t) < \alpha_i \times R_i$). The agent stays idle (no bid
and no ask) with probability 50\%; otherwise it either buys a
share (probability 40\%) or sells a share (probability 10\%).\\

The price offer is again drawn from distribution $p(t-1)+U_{[-1\%;+1\%]}$.

\item[Panic] The risk is higher than the fool's risk ($r(i,t) > \alpha \times R_i$).
The agent preferably sells (probability 90\%), otherwise it stays idle or buys
a share with equal probability (5\%). The price offer is here drawn from
distribution $p(t-1)+U_{[-5\%;0]}$ to account for the panic effects.
\end{description}

To summarize, each agent involves 4 internal parameters: the weights $v_i$
(resp. $w_i$) of the distance to the fundamental value (resp. the price slope)
in the risk function; the agent risk threshold $R_i$; and the fool factor
$\alpha_i$.

After these parameters and depending on the price history, the agent is
associated a strategy (exuberant, comfort or panic), which governs its
propensity to buy or sell and its price offer.

\section{Experimental results}
\label{Expe}
All the results reported below are based on experiments involving 10 agents
trading for 1000 iterations. By default, agents have $R_0=0.5$, $F_i(t)=100$,
$v=1$, $w=-3$, $\alpha=1.1$. They are additionally endowed with 1000 units of
cash and a random number of shares comprised between 0 and 10.


    \subsection{Exogenous risk}
\begin{figure}
            \centering

                \includegraphics{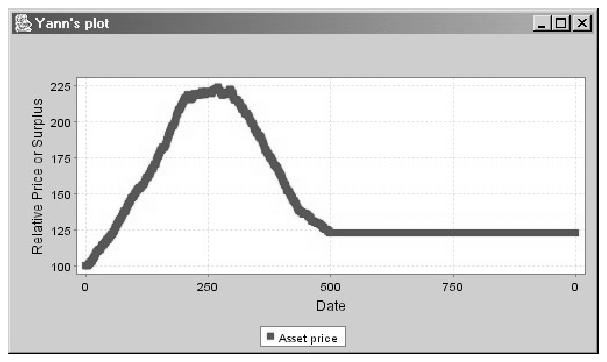}

                \includegraphics{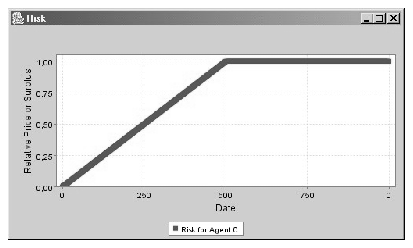}
                \includegraphics{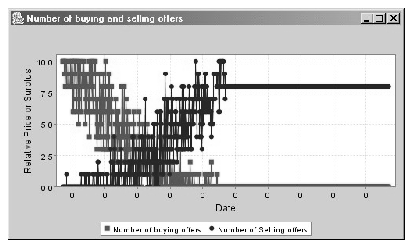}

                \caption{(left) Risk is set exogenously to simulate linearly decreasing
                hopes for dividend collecting.
                (right) Corresponding numbers of buying and selling offers
                (center) A speculative bubble is observed
                wrt a fundamental value of 100
                }
            \label{fig:linearBubble}
            \end{figure}

    With the first model of rationality, we obtain, as expected, a very clean
    bubble. The simplest case, illustrated by figure \ref{fig:linearBubble},
    shows a linear and symmetric bubble that corresponds to a strictly linear
    risk estimation. Prices are climbing while $r(t)$ remains below 0.5 and
    then start to go downward as selling shares becomes more likely than buying
    shares. More realistic looking bubbles with a steeper crash were obtained by using
    an $\arctan$-based function instead of a linear one but as the behaviour
    observed in this case is trivially dictated by the shape of the risk
    function, these experiments should not get much more attention nor be seen
    as anything else than an empirical proof of consistence for our
    implementation.

    \subsection{Endogenous risk}
    The second case is much more interesting. Market behaviour is not trivial
    anymore with respect to the risk estimation function and we are going to see how
    one can navigate between efficiency and ``bubbly behaviour'' by playing on the distribution
    of parameter values across the population.

        \subsubsection{Efficiency}
\begin{figure}
                \centering
                \includegraphics{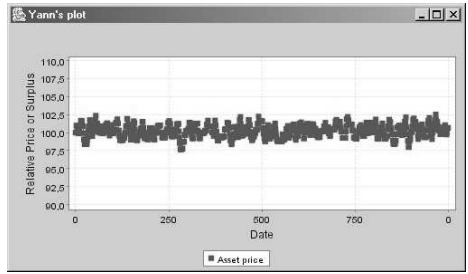}

                \caption{When there is no heterogeneity and prices are close to $F$, the market is stable: prices remain within 2 or 3 percent of F}
                \label{fig:efficiency100}
            \end{figure}

\begin{figure}
                \centering
                \includegraphics{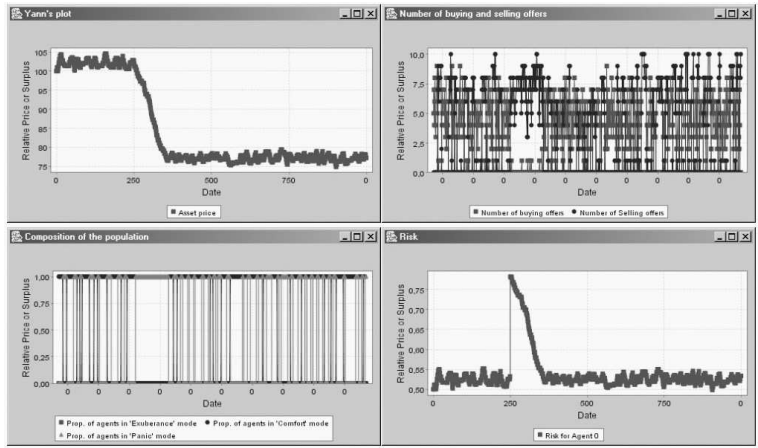}

                \caption{The market is nearly efficient: at iteration 250, F is switched to 75 and prices quickly reflect this sudden change in common beliefs.}
                \label{fig:efficiency75}
            \end{figure}

        When the population of agents is initialized homogenously with the default values, 
        the market is, as illustrated by figure \ref{fig:efficiency100}, relatively
        stable as prices remain close (within a 2\% or 3\% range) of $F$. As illustrated
        by figure \ref{fig:efficiency75}, when $F$'s value is suddenly changed, prices quickly
        reflect this modification. The market appears in such cases and
        naturally enough, as nearly efficient as it reflects stability or changes in common
        beliefs.

        \subsubsection{``Bubbly'' behaviour}

\begin{figure}
                    \centering
                    \includegraphics{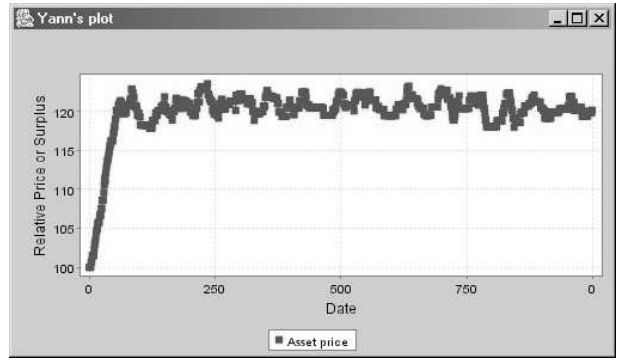}

                    \caption{A speculative bubble with no crash obtained by distributing $R_0$ in [0.4;0.8]}
                    \label{fig:bubbleNoCrash}
                \end{figure}

            \paragraph{Bubbles without a crash}
                All other parameters being the same for all agents, as soon as one
            introduces heterogeneity in risk tolerance ($R_0$), which means agents
            are initialized with a value for $R_0$ that is drawn from a uniform
            distribution instead of being the same for everyone, speculative
            behaviour start to appear with prices that tend to move away from the
            commonly held view of $F$'s value. For instance, as shown in figure
            \ref{fig:bubbleNoCrash}, with a biased upward spread for $R_0$ of 0.4, which
            means that agents have values for $R_0$ that lie between 0.4 and 0.8
            instead of having a common default value of 0.5, we observe a rise in prices
            followed by a jittery oscillation.

\begin{figure}
                \centering

                    \includegraphics{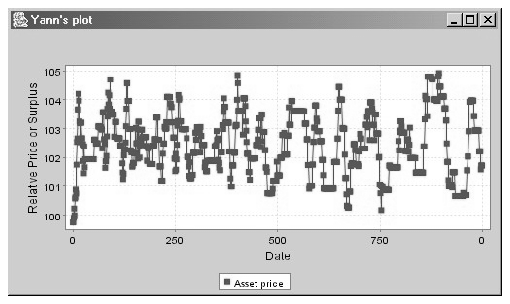}
                    \includegraphics{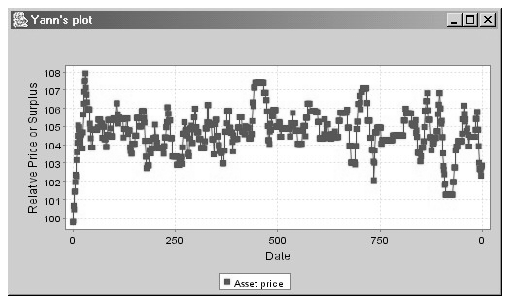}

                    \includegraphics{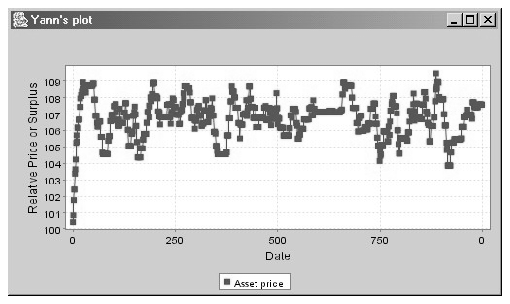}
                    \includegraphics{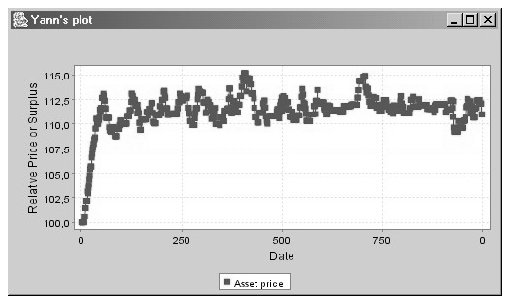}

                    \caption{Different values for $\alpha$: (top left)1.0,(top right)1.05,(bottom left)1.1,(bottom
                    right)1.2. Increasing $\alpha$ increases the
                    magnitude of the oscillation.}
                \label{fig:foolFactor}
                \end{figure}

            \paragraph{Influence of $\alpha$}
            This is conditioned by the ``fool factor'''s value $\alpha$. Figure
            \ref{fig:foolFactor} show that the bigger alpha is, the larger the
            speculation's magnitude gets. This seems natural as increasing
            $\alpha$ increases the number of agents that, under similar risk
            conditions, keep on buying shares a large proportion of the time.

\begin{figure}[h]
                    \centering
                    \includegraphics{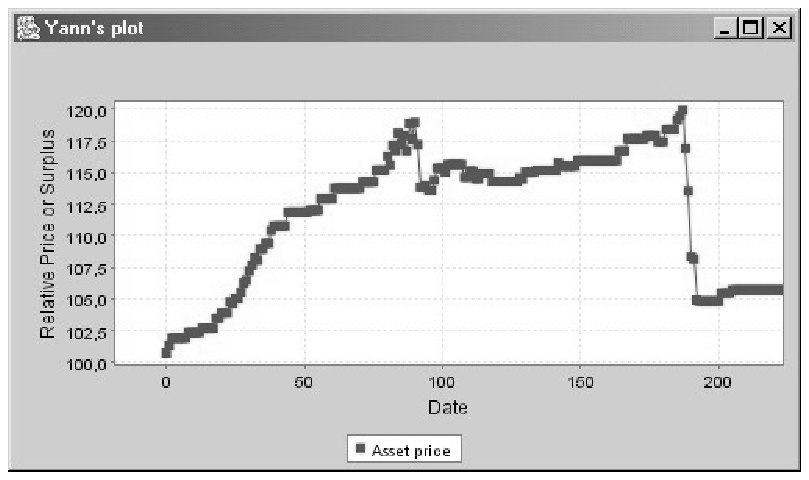}

                    \caption{A speculative bubble obtained with $R_0$ in $U_{[0.4;0.8]}$, $w=-5$ and asymmetric pricing when agents are panicking}
                    \label{fig:bubble}
                \end{figure}

            \paragraph{Introducing asymmetry}
            Real speculative bubbles (i.e. price explosions followed by a steep crash) can only
            be obtained when, additionally to biased heterogeneity in $R_0$, one
            introduces asymmetry in the agents' pricing strategy for the panic
            mode (they use $U_{[-5\%;0]}$ instead of $U_{[-1\%;+1\%]}$). A slight increase in the sensitivity to the price's derivative
            ($w=-5$instead of $w=-3$) also helps to obtain full and sudden
            crashes. Figure \ref{fig:bubble} shows such a bubble.

\section{Conclusion}

We described a simple computational simulation of a financial asset trading
environment where a community of artificial agents base their decisions to buy
or sell shares on an estimation of current market risk. Two contexts were
studied for this estimation: a restricted one with a finite time horizon and an
exogenous risk function and a more refined one, free of any time limitation and
based on a more sophisticated, although still bounded, rationality for
individual agents. Within this restricted model, we experimentally derive a set
of necessary and sufficient conditions for the existence of speculative
phenomena, that is variations in prices that are not explainable by the asset's
underlying fundamental value, typically upward bubbles and sudden crashes.
These conditions are somewhat fuzzy and correlated but clearly state the
importance of heterogeneity, asymmetric behaviour and sensitivity to recent
trends in the birth and rise of financial panics.
\bibliographystyle{plain}
\bibliography{semetCF04}

\end{document}